\documentclass[12pt]{iopart}
\usepackage{bbold}
\usepackage{bm}
\usepackage{graphicx} 
\usepackage[dvipsnames]{xcolor}    
\usepackage{hyperref}
\usepackage{textcomp}
\usepackage{amssymb} 
\usepackage{wasysym}
\usepackage{epstopdf}
\usepackage{cite}

\begin{document}

\title{On the Cutting Edge: Helical Liquids in Time-Reversal-Invariant Topological Materials}

\author{Chen-Hsuan Hsu$^{1}$\footnote{Corresponding author: chenhsuan@as.edu.tw}, Jelena Klinovaja$^{2}$ \& Daniel Loss$^{2}$ } 

\address{$^{1}$Institute of Physics, Academia Sinica, Taipei 11529, Taiwan} 
\address{$^{2}$Department of Physics, University of Basel, Klingelbergstrasse 82, CH-4056 Basel, Switzerland  }

\vspace{10pt}
\begin{indented}
\item[] \today
\end{indented}

\begin{abstract}
In this perspective, we discuss the unique electronic properties of helical liquids appearing at the boundaries of time-reversal-invariant topological materials and highlight the key challenges impeding progress in this field. We advocate for a deeper theoretical understanding of the many-body aspects of these systems to gain insights into helical liquids and the potential stabilization of topological zero modes. Such advancements are crucial for extensively exploring quantum phenomena and for the advancement of quantum science and engineering.

\end{abstract}

{\it Introduction.}  
In the past two decades, time-reversal-invariant topological materials have emerged as an intriguing area of research.
 The theoretical proposal of quantum spin Hall insulators (QSHIs) intimately connects condensed matter research with the concept of topology that extends beyond the conventional Landau paradigm~\cite{Kane:2005a,Kane:2005b,Moore:2007}. 
This breakthrough enabled the prediction of QSHIs in realistic material systems~\cite{Bernevig:2006,Konig:2007,Liu:2008,Knez:2011,Cazalilla:2014,Qian:2014,Wu:2018,Reis:2017,Collins:2018} and laid the foundation for the discovery of diverse topological phases of matter~\cite{Hasan:2010,Qi:2011,Weber:2024}.
   The recent experimental development of twisted bilayer structures~\cite{Cao:2018a,Cao:2018b} has also advanced the search for time-reversal-invariant topological states of matter, leading to recent observations of double or triple QSHI states in twisted bilayer MoTe$_2$~\cite{Kang:2024a} and WSe$_2$~\cite{Kang:2024b}.

While topological band theory in the single-particle description has been proven successful under common circumstances, incorporating many-body effects can reveal intriguing aspects.
For instance, in two-dimensional materials such as transition metal dichalcogenides, incompletely screened Coulomb interactions can result in the formation of strongly bound electron-hole pairs, also known as excitons~\cite{Mueller:2018}. In monolayer WTe$_2$, such phenomena allow for a rich platform for the interplay between non-trivial topology and Coulomb interactions~\cite{Jia:2021,Sun:2022}. 
Remarkably, monolayer TaIrTe$_{4}$ has recently been identified as a dual QSHI, consisting of a single-particle QSHI state at the charge neutrality point and a correlated QSHI state with certain dopings~\cite{Tang:2024}.   
These developments establish the foundation for exploring correlated and topological phenomena in low dimensions.

\begin{figure}[t]
    \centering
 \includegraphics[width=0.5\linewidth]{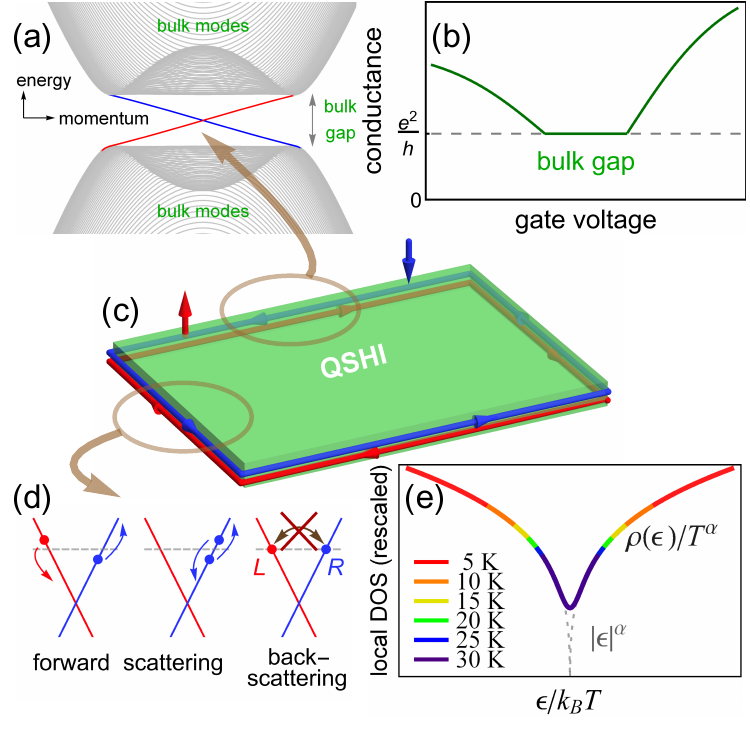}
\caption{Fundamental properties and observable features of helical liquids, spanning their band structure, spectroscopic signatures, and transport behavior.  
(a) Energy spectrum of helical edge states crossing the bulk gap, illustrating spin-momentum locking.  
(b) When the Fermi level, tuned by the gate voltage, lies within the bulk gap, the edge conductance becomes quantized at the universal value $e^{2}/h$ (with $e$ the elementary charge and $h$ the Planck constant) in the absence of elastic backscattering, as illustrated in panel~(d).  
(c) Schematic illustration of a QSHI protected by time-reversal symmetry. The gapless modes reside in the edge channels, with counter-propagating spin states indicated in red and blue.  
(d) In helical liquids, elastic backscattering processes are forbidden in the absence of spin-flip mechanisms, while forward scattering due to electron-electron interactions remains allowed and can lead to correlated effects.  
(e) The rescaled local density of states $\rho(\epsilon)$ follows the curve described by Eq.~(\ref{Eq:DOS}), where the exponent $\alpha$ depends on the forward-scattering strength of electron–electron interactions, as schematically depicted in panel~(d). 
}
    \label{fig:Fig1}
\end{figure}

{\it Helical liquids and their features.} 
Beyond the bulk topology, the boundary states of time-reversal-invariant systems provide fertile ground for realizing correlated one-dimensional quantum matter. 
Another feature arising from many-body effects manifests at the edges of QSHIs or the hinges of higher-order topological insulators~\cite{Schindler:2018sa}. As illustrated in Fig.~\ref{fig:Fig1}(a,c), these boundary channels support one-dimensional gapless modes~\cite{Roth:2009,Fei:2017,Tang:2017,Schindler:2018,Hsu:2021}, characterized by electrons with opposite spins moving in the opposite directions. Analogous to nonhelical one-dimensional electronic systems~\cite{Giamarchi:2003}, electronic interactions within these channels can lead to correlation effects, resulting in the formation of {\it helical liquids}~\cite{Hsu:2021,Bouchoule:2025}.

Protected by the electronic topology of their higher-dimensional bulk, the helical liquids represent a novel class of quantum matter in one dimension~\cite{Wu:2006,Xu:2006,Hsu:2021}, characterized by strong correlations, unusual spin textures, and distinctive transport properties. Their existence allows for the exploration of transport and correlation phenomena in low dimensions. 
Specifically, within the edge channels, the electron spin direction is locked to the direction of movement. 
Time-reversal symmetry thus protects the helical edge channels from elastic backscattering, as any backscattering process would require a spin flip; see Fig.~\ref{fig:Fig1}(d). 
This protection leads to the quantization of edge conductance.
Namely, in an ideal helical liquid at zero temperature, the edge conductance is quantized as
\begin{equation}
    G = \frac{e^{2}}{h},
\end{equation}
when the chemical potential lies within the bulk gap, as demonstrated in Fig.~\ref{fig:Fig1}(b). 
On the other hand, when the time-reversal symmetry is broken, for example by magnetic impurities~\cite{Altshuler:2013,Vayrynen:2016,Hsu:2017,Hsu:2018b} 
or external magnetic fields~\cite{Lezmy:2012,Delplace:2012}, 
spin-flip backscattering processes can emerge, leading to deviations from the quantized conductance. 
Therefore, conductance quantization serves as a hallmark of charge transport in helical liquids and provides a direct probe of their underlying topological character. 
Table~\ref{Table:materials} summarizes representative platforms exhibiting helical edge states, comparing their bulk gaps, operating temperature ranges, and degrees of topological robustness.

\begin{table}[bt]
\centering
\renewcommand{\arraystretch}{0.89}
\caption{Representative platforms for realizing helical liquids and their characteristic parameters. 
The listed bulk gaps and temperature ranges are approximate values reported in experiments.}
\begin{tabular}{lp{2cm}p{2.5cm}p{5.5cm}}
\hline
Material system & Bulk gap  & Operation temperature & Remarks  \\
\hline
HgTe/CdTe~\cite{Konig:2007,Roth:2009,Lunczer:2019} & $14$--$55$~meV & $<10$~K & Quantized conductance observed in short edges; sensitive to sample quality and disorder in long edges. \\
InAs/GaSb~\cite{Knez:2011,Suzuki:2013,Irie:2020} & $3$--$35$~meV & $<10$~K & Type-II QSHI with hybridized electron-hole bands; quantized conductance observed in short edges. \\
Monolayer WTe$_2$~\cite{Tang:2017,Wu:2018,Jia:2022} & $\sim 55$~meV & up to $\sim100$~K & Large bulk gap; STM/STS observation of helical liquids. \\
Bismuthene on SiC~\cite{Reis:2017,Stuhler:2019} & $\sim 0.8$~eV & 77~K & Large bulk gap; robust topological protection with STM/STS observation of helical liquids. \\  
Twisted bilayer MoTe$_2$~\cite{Kang:2024a} & $\sim 0.3$~meV & $<10$~K & Tunable topological gaps and correlated states; potential platform for fractional QSHI regime. \\
Twisted bilayer WSe$_2$~\cite{Kang:2024b} & $1.5$--$4$~meV & $<20$~K & Double QSHI protected by Ising spin conservation symmetry. \\
Monolayer TaIrTe$_4$~\cite{Tang:2024} & $10$--20~meV & up to $\sim100$~K & Dual QSHI with coexisting correlated and single-particle topological states at different doping levels. \\
\hline
\end{tabular}
\label{Table:materials}
\end{table}

Meanwhile, unlike higher dimensions, interacting electrons in the one-dimensional edges cannot be represented as quasiparticles within the Fermi liquid theory. 
The resulting strong correlations produce discernible spectroscopic features that are observable through scanning tunneling spectroscopy (STS).
As displayed in Fig.~\ref{fig:Fig1}(e), the rescaled local density of states at different temperatures follows a single curve given by~\cite{Hsu:2021,Stuhler:2019,Jia:2022}
\begin{eqnarray}
\frac{ \rho_{\rm  } (\epsilon ) } {T^{\alpha_{\rm  }} }
 &\propto & \cosh\left( \frac{ \epsilon }{2k_{\rm B} T} \right)  
 \left| \Gamma \left( \frac{ 1 + \alpha_{\rm  }}{2} +  \frac{ i \epsilon  }{2 \pi k_{\rm B} T} \right)\right|^2,
 \label{Eq:DOS} 
\end{eqnarray}
with a scaling exponent $\alpha$ depending on the interaction strength, the energy $\epsilon$ from the Fermi level, the temperature $T$ and the Boltzmann constant $k_{\rm B}$. 
Equation~(\ref{Eq:DOS}) provides a representative expression for the temperature- and energy-dependent spectral behavior of helical liquids. The scaling exponent $\alpha = (K + 1/K - 2)/2$ reflects the strength of electron-electron interactions through the interaction parameter $K$, which can be tuned in nanoscale systems by controlling the screening effect~\cite{Stepanov:2020,Wang:2024}. Stronger repulsion interaction strength leads to a more pronounced suppression of the low-energy density of states.
The spectroscopic features have been observed in STS  measurements on bismuthene~\cite{Stuhler:2019} and WTe$_2$~\cite{Jia:2022}. Interestingly, these measurements indicate intermediate to strong interaction strengths due to strong confinement of the one-dimensional channels, suggesting a suitable platform for realizing correlated quantum systems~\cite{Hsu:2021}.

In addition to the spectroscopic features, many-body interactions in helical liquids also manifest in transport properties, 
such as  interaction-driven backscattering effects and power-law conductance corrections. 
These behaviors have been extensively analyzed in recent studies~\cite{Lezmy:2012,Altshuler:2013,Vayrynen:2016,Hsu:2017,Hsu:2018b,Hsu:2021}, highlighting that helical liquid characteristics can be studied beyond the local density of states. 
Together, these observations capture the essential physics of helical liquids and set the stage for understanding their potential and future research directions.

{\it Current challenges and open questions.}
Despite significant progress, open questions and challenges persist in the research of topological materials hosting helical liquids. One challenge is consistently achieving conductance quantization in helical channels, which is essential for low-dissipation nanoelectronics that significantly reduce power consumption. Detailed analyses have shown that helical liquids are susceptible to various backscattering mechanisms, which arise from the breaking of time-reversal symmetry or inelastic processes in realistic settings~\cite{Hsu:2021}. Here, again, electronic interactions play an important role in these mechanisms. This issue becomes critical when attempting to scale up these systems for broader applications, as the detrimental effects of disorder or noise may also scale with the system size~\cite{Hsu:2021}.
From the experimental side, potential solutions involve using strain in fabrication to enlarge the bulk gap~\cite{Irie:2020} and applying sweeping gate voltages to shape the potential landscape during measurements~\cite{Lunczer:2019}.
While these techniques yield promising results for QSHIs formed in semiconductor quantum well structures, consistently achieving conductance quantization in two-dimensional materials would be advantageous. The latter systems, including those reported in Refs.~\cite{Fei:2017,Wu:2018,Kang:2024a,Tang:2024}, can be more easily integrated with other layered materials for device engineering. 
It is also crucial to investigate topological phase transitions that occur upon adjusting system parameters, such as external electric and magnetic fields, pressure, or temperature. These studies not only enhance the stability of these systems but also provide new routes for controlling helical liquids in practical applications. Furthermore, exploring new materials and mechanisms is beneficial, particularly for identifying materials with larger bulk gaps to enhance topological protection. In this context, employing advanced theoretical techniques such as quantum chemistry or machine learning to predict and classify topological materials\cite{Vergniory:2019} can play a crucial role in advancing the field.
Overcoming these limitations will not only enhance material robustness but also enable the design of new device architectures, as discussed next.

\begin{figure}[t]
    \centering
 \includegraphics[width=0.5\linewidth]{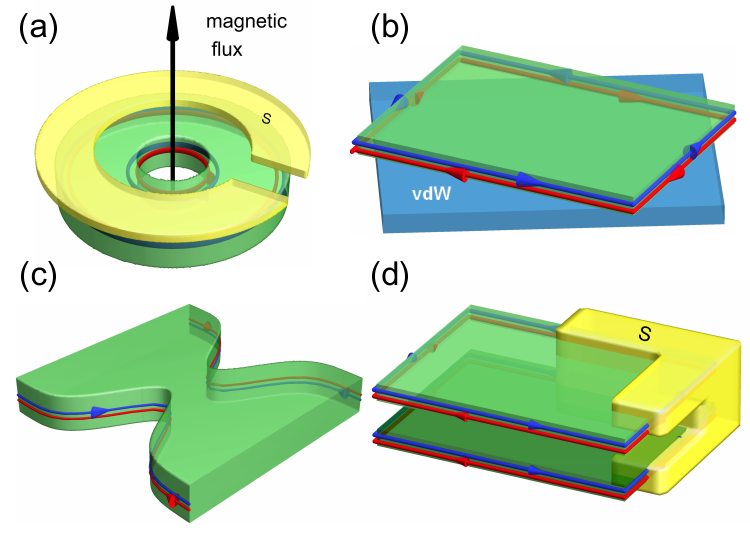}
\caption{Illustrations of device concepts for probing helical liquids or exploiting them in realistic architectures. 
(a) A Josephson junction composed of a QSHI ring and a C-shaped superconducting layer, where the supercurrent is mediated by helical edge states and its flux dependence reveals the topological nature of the junction. 
(b) Heterostructure integrating a QSHI layer with a van der Waals (vdW) material, allowing proximity-induced spin-orbit coupling or superconductivity. 
(c) Quantum point contact formed in a QSHI, serving as a controllable constriction for probing backscattering and correlation effects when multiple helical edges are brought into proximity. 
(d) Two QSHI layers coupled to a superconductor, where nonlocal pairing can stabilize Majorana Kramers pairs, and in the fractional regime, parafermion zero modes may emerge.
}
    \label{fig:Fig2}
\end{figure}

{\it Potential applications and future directions.}
Another challenge involves leveraging helical liquids for applications such as spintronics and robust quantum computing platforms.
To achieve this, one can integrate the system with existing technologies and design compatible devices, as shown in Fig.~\ref{fig:Fig2}. An example is to explore the interplay between helical edges and controllable experimental knobs, such as magnetic fluxes. As illustrated in Fig.~\ref{fig:Fig2}(a), depositing a superconducting layer on a QSHI ring to form a junction~\cite{Fu:2009} enables the measurement of the Josephson current while adjusting the magnetic flux through the ring. In addition to identifying the topological nature through the flux dependence of the supercurrent, extending the setup can also be useful for exploring Josephson effect phenomena under more general circumstances. 
To further investigate helical liquids, heterostructures that combine QSHIs with other layered materials like transition metal dichalcogenides can be used to induce spin-orbit coupling, nontrivial spin textures, or superconductivity~\cite{Guo:2025,Xia:2025}, as illustrated in Fig.~\ref{fig:Fig2}(b). Additionally, quantum point contacts made from QSHIs~\cite{Strunz:2020}, as depicted in Fig.~\ref{fig:Fig2}(c), can be utilized to study transport and correlation phenomena when multiple edges hosting helical liquids are brought into close proximity.

An interesting development occurs when superconductivity is induced in helical edges.  Besides the setup in Fig.~\ref{fig:Fig2}(a), a more complex arrangement, as shown in Fig.~\ref{fig:Fig2}(d), can be employed to stabilize Majorana zero modes without the necessity of external magnetic fields~\cite{Klinovaja:2014}, forming Kramers doublets of topological bound states~\cite{Zhang:2013,Hsu:2018,Haim:2019,Hsu:2024,Hung:2025}.
These zero modes are pivotal for developing topological quantum bits and have applications in quantum computing~\cite{Nayak:2008,Liu:2014,Schrade:2018,Schrade:2022}.  
It was shown that Majorana Kramers doublets in time-reversal-invariant topological superconductors obey non-Abelian statistics protected by time-reversal symmetry and can support topological quantum computation~\cite{Liu:2014}.
Crucially, a prerequisite for the formation of Majorana zero modes is sufficiently strong electron-electron interaction within the helical channels~\cite{Klinovaja:2014,Hsu:2018,Hsu:2024}, a condition demonstrated in the aforementioned spectroscopic measurements on bismuthene~\cite{Stuhler:2019} and WTe$_2$~\cite{Jia:2022}. 
Moreover, many-body interactions might lead to novel bulk phases such as fractional topological insulators~\cite{Levin:2009,Maciejko:2015,Neupert:2015}, potentially resulting in {\it fractional helical liquids} at their edges. 
Intriguingly, operating the setup in Fig.~\ref{fig:Fig2}(d) within the fractional regime could allow the edge states to host parafermion zero modes\cite{Klinovaja:2014}, characterized by their unusual fractional statistics and offering enhanced computational capabilities compared to Majorana zero modes. 
Recent experiments on twisted nanostructures allow for the engineering of quasi-flat bands by varying the twisting structure, thereby enhancing many-body correlation effects. Notably, these efforts have demonstrated the potential to form fractional QSHIs in twisted bilayer MoTe$_2$~\cite{Kang:2024a}, underscoring the promise of such systems as platforms for fractional parafermion zero modes.
 These diverse approaches demonstrate that helical liquids serve as versatile building blocks for both fundamental research and quantum technologies.
 
 {\it Conclusion and outlook.}
The distinctive electronic properties of helical liquids in time-reversal-invariant topological materials provide a platform for advancing our understanding of correlated quantum matter and driving technological innovations. In line with this, Weber~et~al. recently offered insights into two-dimensional topological insulators, outlining a vision and roadmap for future endeavors in this field~\cite{Weber:2024}. Challenges persist in identifying better material platforms, improving the design of hybrid electronic devices, and bridging the gap between expected outcomes from idealized models and observations in realistic settings. Addressing these challenges holds the promise of unlocking the full potential of these materials and establishing a solid foundation for quantum science and sustainable technology. 
Continued synergy between material discovery, device engineering, and theoretical modeling will be essential to unlock their full potential in future quantum technologies.

In outlook, several concrete directions emerge for advancing the study of helical liquids. Promising material platforms include monolayer WTe$_2$, bismuthene, and twisted transition-metal dichalcogenides, 
where strong spin-orbit coupling and enhanced correlations enable robust edge transport. Verification of the helical nature through tunneling spectroscopy and Josephson devices will be crucial for identifying the microscopic origin of topological superconductivity and the formation of zero modes. Developing scalable heterostructures that integrate QSHIs with superconductors or correlated materials will further deepen our understanding of quantum phenomena in low-dimensional systems. On the theoretical side, combining many-body and single-particle approaches, such as bosonization, first-principle calculations and machine-learning-assisted material search, offers a route toward predictive modeling of correlated topological systems. Together, these efforts will provide the experimental and theoretical foundations of helical liquids and guide the realization of next-generation quantum electronic devices.

{\it Acknowledgments. } 
We thank Bent Weber for engaging in stimulating discussions and sharing his insights in this field. 
This work was financially supported by the Swiss National Science Foundation (Switzerland), the NCCR QSIT, and the National Science and Technology Council (NSTC), Taiwan, through Grant No.~NSTC-112-2112-M-001-025-MY3 and Grant No.~NSTC-114-2112-M-001-057, and Academia Sinica (AS), Taiwan through Grant No.~AS-iMATE-114-12.

{\it Data availability statement. } 
No new data were created or analysed in this study.

~ \\ 
 

\providecommand{\newblock}{}

\end{document}